\documentclass[fleqn,usenatbib,useAMS,onecolumn]{mnras}

\usepackage{graphicx}	
\usepackage{amsmath}	
\usepackage{amssymb}	
\usepackage{multicol}        
\usepackage{bm}		
\usepackage{pdflscape}	
\usepackage{changes}

\usepackage{epstopdf}

\usepackage[T1]{fontenc}
\usepackage{ae,aecompl}

\usepackage{newtxtext,newtxmath}

\title{Radiative energy from a reconnection region around massive black hole}

\author[T.L. Zhao et al.]{
	Tian-Le Zhao,$^{1,2}$\thanks{E-mail:tianle@mail.ustc.edu.cn}
	and Rajiv Kumar$^{1,2}$
	\\
	$^{1}$CAS Key Laboratory for Research in Galaxies and Cosmology, Department of Astronomy, University of Science and Technology of China, Hefei 230026, China \\
	$^{2}$School of Astronomy and Space Sciences, University of Science and Technology of China, Hefei 230026, China\\
}
\date{Last updated XX; in original form XX}
\pubyear{2020}
\begin{document}
\label{firstpage}
\pagerange{\pageref{firstpage}--\pageref{lastpage}}
\maketitle
\begin{abstract}
In the previous numerical study, we find the blob formation and ejection in the presence of magnetic reconnection in the environment of the hot flow of the accretion disk. Based on those encouraging results, in the present work, we calculate the energy and the spectrum of the emission in the different bands around Sgr A*. We assume the electrons in the magnetic reconnection region are non-thermal and emits synchrotron radiation. The electrons in the other region are thermal, which follows the thermal distribution, and the thermal electron emission mechanism is thermal synchrotron radiation. During the whole process of the magnetic evolution and reconnection, we find two peaks in the temporal light curve in the recently observed Radio frequencies ($230$~GHz and $43$~GHz) and NIR wavelengths ($3.8\mu m$ and $2.2\mu m$). Although, the light curve of the NIR band is most prominent in a single peak. 
The first peak appears because of the blob in the plasma flow, which is formed due to the magnetic reconnection. The second peak appears due to the production of the non-thermal electrons with the evolution of the magnetic flux. Both peaks reach luminosity more than $10^{26}$~erg/s for a single plasmoid/blob. For the NIR band, the highest luminosity can reach  more than $10^{28}$~erg/s.These luminosities can be high for the large simulation area and the stronger magnetic field with the multiple blobs. We infer that the observed flares are a group of magnetic reconnection phenomena, not a single one.
\end{abstract}



\begingroup
\let\clearpage\relax
\tableofcontents
\endgroup
\newpage

\section{Introduction}

Being the centre of the Milky Way and a compact radio power\citep{2010RvMP...82.3121G}, sagittarius A* (SGR~A*) could be observed by many wavebands, e.g., radio sub-mm \citep{Bower2015,Bower2019,Brinkerink2015,Liu2016}, far infrared\citep{VonFellenberg2018}, near-infrared \citep{2018ApJ...863...15W} and X-ray flux \citep{Baganoff2003}. Their radiation processes are quite persistent which could be referred to the characteristic radiation around the center of the black hole \citep{1999ApJ...524..816R,1999ApJ...524..805B,2003ApJ...598..301Y}. The SGR~A* is found to be a low luminous object with a low accretion rate \citep[and references therein]{2014ARA&A..52..529Y} and they are very well-fitted with the advection-dominated accretion flow (ADAF) model in multi-wavelength spectra from the radio to X- ray\citep{1995Natur.374..623N, 1998ApJ...492..554N, 2003ApJ...598..301Y, 2004ApJ...606..894Y}. \citet{2009ApJ...699..722Y} has calculated the images of the ADAF surrounding a Kerr black hole with the observational wavelengths, with primary analysis by comparison with the observed sizes of SGR~A* suggests that the disk around the central of SGR~A* is highly inclined, or the central black hole is rotating fast. SGR~A* is the most observed supermassive black hole and the most suitable observable object in many instruments such as {\it ESO VLT} (Paranal/Chile) and {\it SINFONI} and {\it VISIR}\citep{2020ApJ...897...28P}. The Event Horizon Telescope {\it Event Horizon Telescope} (EHT){ will soon provide its first high-resolution images\citep{2021arXiv210108618F}.\newline

 Many observations also indicate that there are multiple hot spots and many flare events in the source.  Based on the observation of the SGR A* by Chandra, \citet{2002A&A...383..854Y} proposed different models, and magnetic reconnection is one of the models responsible for the blobs and flares in microphysics of the ADAF.  Recently, {\it Very Large Telescope Interferometer} (VLTI) observed several near-infrared super flares which might be from the hot spots at the inner orbit of the SGR~A* \citep{2018A&A...618L..10G, 2020ApJ...891L..36G}. And it is believed that these hot spots are likely formed by relativistic magnetic reconnection event. \citet{2020arXiv200901859G} studied the polarization data of SGR~A* during a bright NIR flare observed with the {\it GRAVITY}  on July 28, 2018, the beam depolarized like emission indicated the flares' emission region resolves the magnetic field structure near the blackhole. \citet{2020ApJ...897...28P} through the observations of near- and mid-infrared of {\it VISIR}, find a proper motion of the filamentary emission which is very close to SGR~A*. Also, \citet{2020MNRAS.498.4379K} found the bright flares of X-ray emission of SGR~A*,  combining these data with measurements of polarization can indicate that SGR~A* is the primary source and the emission of the flares is polarization properties.  \citet{2020A&A...638A...2G} calculated the data with high resolution of the  {\it VLTI}, using interferometric model fitting and consistent flux measurements to obtain light curves, indicated the emission is generated in a log-normal process and create the observed power-law extension of the flux distribution. Also, \cite{Ponti2017} show a bright simultaneous flare at the NIR and X-ray flux density spectrum that they observed. All these observations indicated that the bright, episodic $X$-ray and near-infrared flares are the common phenomena on the ADAF of SGR~A*. \newline
 
 Time-dependent simulations based on the magnetohydrodynamic (MHD) equations can show the evolution of plasma, such as magnetic reconnection\citep{2020MNRAS.499.1561Z}, and by calculating the radiation energy can  be able to reproduce the typical observational characteristics of the light curves at many wavebands. Many simulations\citep{2014MNRAS.442.2797D,2015ApJ...812..103C,2017MNRAS.467.3604R} result about the accretion flows around the supermassive black hole(SMBH) produced realistic light curves. Some simulations introduce the non-thermal electrons to calculate the light curves during the evolution of the MHD simulation \citep{2016ApJ...826...77B,2017MNRAS.466.4307M}. Some models assuming the light curve of bright NIR/X-ray flares have been a distribution of electrons is accelerated from thermal equilibrium into a power law energy distribution\citep{2010ApJ...725..450D,2015ApJ...810...19L,2017MNRAS.468.2447P}. However, it can't be very well understood the mechanism of this acceleration in a qualitative fashion\citep{2020arXiv200901859G}. As several mechanisms in the previous works,  magnetic reconnection is a possible mechanism in analogy with solar flares or coronal mass ejections\citep{2003ApJ...598..301Y,2006ApJ...650..189Y}. However, all of these simulations are restricted by the size of the simulation area, numerical resolution of the simulation, and the initial magnetic field configuration is uncertain. It will be difficult to produce realistic outflows and cannot produce the observed high energy fluxes during hot-spots and flares\citep{2020arXiv200901859G}. \citet{2020MNRAS.495.1549N} present a series of large scale two-dimensional GRMHD simulations of different kinds of tori accreting on rotating black holes to study the formation and evolution of current sheets. \citet{2021arXiv210501145R} use the high-resolution 3D simulation to study the two counter-propagating Alfv\'{e}n waves, the dissipation mechanism is magnetic reconnection and show the current sheets form as a natural result of nonlinear interactions between counter-propagating
 Alfv\'{e}n waves. \citet{2021MNRAS.502.2023P} use the 3D GRMHD simulation to study the magnetically arrested accretion discs (MADs) and show the magnetic dominant regions can resist being disrupted via magneto-rotational turbulence and shear and the orientation of the magnetic field is predominantly vertical as suggested by the GRAVITY data of the SGR A*. \newline

In our previous work, \citet{2020MNRAS.499.1561Z} set up a small numerical box on the accretion disk around the SGR~A*,  the blobs and flux structure formed during the process of the magnetic reconnection, which proved the theoretical results of \citet{2009MNRAS.395.2183Y} in presence of the hot flow around the accretion disk. Our simulation is in small scales and shows the detail process of the blob and flares formed in the current sheet during the magnetic reconnection, which is main difference from the others work. They have shown the whole disk of the SGR A*, and in many of the works, the process of the magnetic reconnection is just a point on the disk. And the present study has shown the details of the emission and evolution of the blobs and magnetic flux structure with nicely following the magnetic reconnection processes(more details see in \citet{2020MNRAS.499.1561Z}). In the present study, our goal is to study the radiative  properties from those magnetic reconnection sites, we can give us some understanding of the recent observations of the SGR~A*. Here we consider the thermal electrons and non-thermal electrons with their corresponding radiative emissivities in our model. The magnetic reconnection can provide both thermal electrons and non-thermal electrons due to magnetic dissipation. 
We consider that the non-thermal electron can be found in the regions of the blob formation and magnetic flux evolution, which can provide non-thermal synchrotron radiations. 
In next section \ref{sec2}, we show the calculation of the radiative mechanisms. In section \ref{sec3}, we show the results. In final section \ref{sec4}, we conclude  our studies.\newline

\section{Calculation of radiative emission}
\label{sec2}

As described in \citet{2020MNRAS.499.1561Z}, the 2D model is set up with the single-fluid MHD equations. The initial condition is the hydrostatic equilibrium of the black hole accretion disk environment. It is a small simulation area on the surface of the accretion disk. It can be approximated to a point on the accretion disk. The outflow boundary condition is set on left-hand, right-hand and upper boundaries.  Two ghost layers below the physical is inserted in bottom boundary. The magnetic field inside the two layers with the ghost grid cells is set as the formula that described in \citet{2017ApJ...841...27N,2018RAA....18...45Z,2020MNRAS.499.1561Z} that we show in \ref{MHD}. The strength of the magnetic field below the bottom boundary will vary with time.  We calculate the radiation of all points in the numerical simulation area that described in \citet{2020MNRAS.499.1561Z} to show the emission energy change for whole magnetic reconnection process, including blob formation and ejection and flux formation. In the magnetic reconnection region, the density and the temperature of the plasma are higher, the magnetic pressure is stronger. The plasma density around the reconnection region is lower than in the middle of the magnetic reconnection region. Moreover, we have considered only synchrotron radiation of the electron. In the magnetic reconnection region, we assume the mechanism of the electron is synchrotron radiation, and in other areas, we assume the mechanism of the electron is thermal synchrotron radiation.

\subsection{A brief introduction to MHD numerical simulation}
		\label{MHD}
For the initial condition, the simulation area is located at $R_\mathrm{in}=10R_\mathrm{s}$ from the center of the blackhole. The size of simulation box is $300L_0 \times 300L_0 $ (where $ L_0=10^6 $~m) on the surface of the ADAF.  The strength of an initial magnetic field is used to be  $B_{x0} = - 0.6b_0$ and $ B_{y0} = - 0.8b_0$, here $b_0=16G$. The distribution of the gas density  $\rho=2\times10^{-10}$~$(kg m^{-3})$, the gas pressure ($P_0$) and thermal energy density ($E_{t0}$) on left boundary of the box are follows:
		\begin{equation}
			\rho=\rho_{0}\times \exp\left(-\frac{{(x*\sin\theta)}^2}{{H_0}^2}\right)
			\label{rho},
		\end{equation}
		where, $\sin\theta=H_0/R_\mathrm{in}$, and
		\begin{equation}
			P_0=\frac{2\rho_0 k_B T_0}{m_i} , 
		\end{equation}
		
		\begin{equation}
			E_{t0}=\frac{P_0}{\gamma-1}, 
		\end{equation}
		where $T_0=6\times10^{9}$~K is  the gas temperature of the Left most side of the simulation box, and adiabatic index, $\gamma=5/3$, so the distribution of the gas thermal energy can be derived from hydrostatic equilibrium condition as:
		\begin{equation}
			\rho\nabla\Phi=-\nabla P.
		\end{equation}
		\newline
		
		Two extra layers with the ghost cell are set at the each boundary. The outflow boundary conditions are set on left, right and upper boundaries for more details see our previous works \cite{2020MNRAS.499.1561Z,2018RAA....18...45Z}. Two ghost layers are inserted in bottom boundary at $ y = H_0 $. The variation of magnetic field is set inside the two layers with the ghost grid cells as follows:
		\begin{equation}
			b_{xb}=-0.6b_0+\frac{100L_0(y-y_0)b_1 f}{[(x-x_0)^2+(y-y_0)^2]}
			\left[\mathrm{tanh}\left(\frac{x-170L_0}{\lambda}\right)-\mathrm{tanh}\left(\frac{x-230L_0}{\lambda}\right)\right]    , 
		\end{equation}
		\begin{equation}
			b_{yb}=-0.8b_0-\frac{100L_0(x-x_0)b_1 f}{[(x-x_0)^2+(y-y_0)^2]}
			\left[\mathrm{tanh}\left(\frac{x-170L_0}{\lambda}\right)-\mathrm{tanh}\left(\frac{x-230L_0}{\lambda}\right)\right]    , 
		\end{equation}
		\noindent where $t\le t_{1}$ for $f=t/t_{1}$ and $t\ge t_{1}$ for $f=1$,$x_0=R_\mathrm{in}+200L_0$ and $y_0=H_0+6L_0$, $\lambda=0.5L_0$ , $ b_1=3.2\times10^{-4} $, $H_0$ is height of the surface of the ADAF. Here $ t_1=300 $~s. The magnetic field strength below the bottom boundary varies with time till $ t<t_1 $, and it stops after $ t=t_1 $. \newline

\subsection{Non thermal Synchrotron radiation}
In the blobs and magnetic flux structure region, electron distribution of the reconnection region(X-point) can well match the power-law distribution just as the previous research\citep{2001ApJ...562L..63Z,2016MNRAS.462...48S}, and the synchrotron emission is dominating by the non-thermal electrons. Thus the power of the synchrotron per unit frequency emitted by each electron can be written as
\begin{equation}
	\dfrac{dP(\nu)}{d\nu}=\dfrac{2\pi\sqrt{3}e^{2}\nu_{\mathrm{L}}}{c}\left[\frac{\nu}{\nu_{\mathrm{c}}}\int_{\frac{\nu}{\nu_{\mathrm{c}}}}^{\infty}K_\frac{5}{3}(t)dt \right]\mathrm{(erg\cdotp s^{-1}\cdotp {\mathrm{Hz^{-1}}})}, 
	\label{1}
\end{equation}
here, $\nu_{\mathrm{L}}$ is Lamer frequency  $\nu_{\mathrm{L}}=\frac{1}{2\pi}\frac{eB}{m_{0}c}$, and  $\nu_c=\frac{3}{2}\nu_{\mathrm{L}}\gamma^{2}$. $K_\frac{5}{3}$ is a 5/3th order modified Bessel function. The spectra shape of single electron synchrotron spectra is definite by the dimensionless synchrotron radiation spectrum:
\begin{equation}
F\left( \frac{\nu}{\nu_{\mathrm{c}}}\right) =\frac{\nu}{\nu_{\mathrm{c}}}\int_{\frac{\nu}{\nu_{\mathrm{c}}}}^{\infty}K_\frac{5}{3}(t)dt.
\end{equation}
So the  emitting coefficient  of Synchrotron radiation is 
\begin{equation}
	j(\nu)=\int N(\gamma, \alpha,t)d\gamma d\varOmega_{a}\dfrac{dP(\nu)}{d\nu}
\end{equation}
here$N(\gamma, \alpha,t)d\gamma$ is power law electron energy distribution, and $d\varOmega_{a}=2\pi \sin\alpha d\alpha$. \citet{2020A&A...638A...2G} find the flares can create the observed power law extension of the flux distribution\citep{2021arXiv210108618F}, so we assume a one-zone model with a power law electron energy distribution $\gamma_1 \leq \gamma \leq\gamma_2$:
\begin{equation}
	N(\gamma,\alpha)=N_{\gamma}\cdotp\gamma^{-n}g(\alpha)/4\pi,
\end{equation}
here we consider $\gamma_{1}=1$ and $\gamma_{2}=1000$ for most of the figures unless stated.  
As the described in \citet{2018ApJ...862...80B}, the electron power law steepens significantly, and the electron spectrum eventually approaches a Maxwellian distribution, for all values of $\sigma$. In our simulation, the plasma $\beta$ in the reconnection region is mostly around one or greater than one. So, we have chosen parameter, n=1.5.\newline

The luminosity in the jet frame is given by
\begin{equation}
    L_{\nu}=j(\nu)\int\int\int dxdydz,
\end{equation}
Which is the same like \citet{2017ApJ...842..129C}.The Thomson optical depth of the plasma $\tau=n_e\sigma_{T}R $ is less than $1$ in whole simulation, because of our simulation area is too small, so we do not need consider the self-absorption. Consider the electron projection angle $\alpha$ is isotropic distribution, so $g(\alpha)=1$.

\subsection{Thermal synchrotron radiation}
Different with the non-thermal synchrotron radiation, the thermal synchrotron radiation depends on the plasma temperature. The angle-independent emission coefficient of the thermal synchrotron,$j_{\nu}$, of plasma temperature T in the magnetic field, $B$ can be written as \citet{1996ApJ...465..327M} :
\begin{equation}
	j_{\nu}=\dfrac{2^{1/6}\pi^{3/2}e^{2}n_{e}\nu}{3^{5/6}cK_{2}(1/\Theta)\vartheta^{1/6}}a(\Theta,\vartheta)\exp[-(\frac{9\vartheta}{2})^{(1/3)}],
	\label{6}
\end{equation}
here $\vartheta\equiv\nu/\nu_{l}\Theta^{2}$, the correction factor, $a$, can obtained by integration of equation(31-33) in \citet{1996ApJ...465..327M}, $n_e$ is the electron density, and dimensionless plasma temperature $\Theta\equiv kT/m_{e}c^{2}$.
$K_2$ is a modified Bessel function that given by:
\begin{equation}
	K_{2}(\frac{1}{\Theta})\approx\left\lbrace_{2\Theta^{2}-\frac{1}{2}+\frac{ln(2\Theta)+3/4-\gamma_{E}}{8\Theta^{2}}+\frac{ln(2\Theta)+0.95}{96\Theta^4},\Theta \geq 0.65}^{(\frac{\pi\Theta}{2})^{(1/2)}(1+\frac{15\Theta}{8}+\frac{105\Theta^2}{128}-0.203\Theta^3)\exp(-1/\Theta), \Theta\leq0.65;} \right. 
\end{equation}
here $\gamma_{E}\approx0.5772$ is Euler's constant. \newline 

Although thermal synchrotron emission of the optically thin is rarely observable. For relevancy of the optically thin spectra, we need to determine the turnover frequency $\nu_{t}$. If the photon frequency, $\nu<\nu_t$ then the plasma can become optically thick\citep{2000MNRAS.314..183W}.  So the observed spectrum can be dominated by the blackbody radiation, typically in the Rayleigh-Jeans limit (low energy photon, $h\nu<<kT$).
When the photon frequency is greater than the turnover frequency $\nu_{t}$ then the plasma medium can become optically thin for the synchrotron radiation. So we calculate the turnover frequency $\nu_{t}$ as follows and the formulas are given by \citet{1998MNRAS.301..435Z}:
\begin{equation}
	\dfrac{\nu_{t}}{\Theta^{2}\nu_{l}}=\frac{343}{36}ln^{3}\dfrac{C}{ln\frac{C}{ln\frac{C}{\dots}}},     C=\dfrac{3}{7\Theta}	\left[\dfrac{\pi\tau_{_{T}}a~exp(1/\Theta)}{3\alpha_{f}\chi_{c}} \right]^{\frac{2}{7}} 
\end{equation}
here dimensionless cyclotron frequency $\chi_{c}\equiv h\nu_{l}/m_{e}c^2$, $\tau_{_{T}}=n_{e}\sigma_{_{T}}R$ is the plasma Thomson optical depth, $\alpha_f=7.2973506e-3$ is the fine-structure constant, $R=3\times10^{10}$~$cm$ is the characteristic size of the plasma. 
We find maximum turnover frequency, $\nu_{tm}\sim10^{11}$ during the whole simulation as shown in the fig.\ref{Fign0}, which lies in the radio band only. So we calculate the optically thin spectra of the thermal synchrotron radiation for $\nu\gtrsim\nu_{tm}$, which is shown in fig. \ref{Fig00} for $\nu=230GHz$. We find that the thermal luminosity is around several order of magnitude lower than the non-thermal luminosity for the same frequency. Here we calculate the thermal luminosity for the whole simulation box, so this radiation can be considered as the background radiation over the non-thermal emission (Fig. \ref{Fig2}). Moreover, the luminosity of the thermal radiation decreases with increase in the photon frequency, so it becomes very-very low in the NIR band than the non-thermal emission.
We also checked blackbody radiation for $\nu<\nu_{tm}$, which is even lower than the thermal synchrotron luminosity, so we ignore it in our study. 
Thus our main radiation mechanism in present study is the non-thermal synchrotron, which is used in the understanding of the some recent observations of the SGR~A*.

\section{Results}
\label{sec3}

Magnetic reconnection plays an important role during blob formation and ejection.  The aim of the previous work, \citet{2020MNRAS.499.1561Z}  was to understand the magnetic reconnection event and its behaviour with different radii around the black hole. In the present paper, we aim to study the characteristics of the radiative from the blobs and magnetic flux structures, then we can understand some observations around the SGR A*.  In the medium of the magnetized flows, it is very difficult to separate the thermal and non-thermal electrons, since magnetic reconnection can produce an acceleration in the particles and the conversion of the magnetic energy can be heating the plasma. In the previous paper, the particles are accelerated upward in the magnetic reconnection region, so we assumed that the same region can also be dominated by the non-thermal particles.  The magnetic reconnection can provide the current and form a current sheet\citep{2021ApJ...912..109K}. Becuse of the magnetic energy is converted into heat and kinetic energy, the structure of the current sheet(include the outflow region above the current sheet) has higher temperature and lower magnetic strength, so it also has an obvious feature on the distribution of the plasma $\beta$. So to quantify the thermal and non-thermal electrons, we use the plasma $\beta$ parameter (the ratio of gas to magnetic pressure). If $\beta\ge 0.95$ then the electrons are the non-thermal electrons and the rest are the thermal electrons. The distribution of the electron number density and $\beta$ is shown in Figure \ref{Fig0}, and the magnetic reconnection region has a higher $\beta$.  \newline

The simulated light curves of the emission of non-thermal synchrotron with emission of the thermal synchrotron radiation of the background are presented in Figure \ref{Fig2} with two radio frequencies, $230$~GHz and $43$~GHz \citep{2003IAUJD..18E..50M}.  Before magnetic reconnection happens, electrons are all thermal electrons, and the energy of their thermal synchrotron radiation is around three orders of magnitude lower than that of the peak luminosity of the synchrotron emission. As the magnetic process started in the simulation box, the number of the thermal electron decreases and the number of the non-thermal electron increases, this process make the energy of the emission of the synchrotron increased and form the first peak ($t\sim250$~s) as shown in the light curve. At the same time the average magnetic field strength decreases and becomes lowest at the time $t\sim300$~s (Figure \ref{Fig1}), so the emission of the synchrotron also decreased (Figure \ref{Fig2}). When the reconnection process becomes weak (reduces magnetic energy dissipation/conversion) and the blob has been ejected, the magnetic field raises again with peaking around $t\sim400$~s. Around $t\sim 400s$, the magnetic flux structure is starting to evolve with reconnection between pre-existing and emerging field lines, the  non-thermal electrons become more than the beginning of the magnetic reconnection. Thus, as shown in the last panel of Figure \ref{Fig0}, due to the persistent nature of the magnetic flux structure, the second peak in the light curve formed by the process of magnetic flux formation, which is wider than the first peak. Two peaks in the light curve are formed by the blob/plasmoid and the flux structure, which could be similar to the July 28th, 2018 NIR {\it GRAVITY} observation \citep{2020arXiv200901859G}. Although the peak separation time, frequency, and luminosity in our work are quite different from the {\it GRAVITY} observation, the process of the magnetic reconnection in \citet{2020MNRAS.499.1561Z} could be one of the reasons for the formation of the double peak light curve in higher frequency around the SGR A*\citep{2020arXiv200514251B}. \newline

Figure \ref{Figbb} show the simulated light curves of the emission of synchrotron with emission of the thermal synchrotron emission of the background with in both wavebands $\lambda=3.8$~$\mathrm{\mu m}$ and $\lambda=2.2$~$\mathrm{\mu m}$. At first, the  emission energy is almost zero, the non-thermal electrons is little at this time. After $t=200$~s, the first peak appears, then the emission energy decrease, because the magnetic field is decreasing (Figure \ref{Fig1}), and the blob is moving out (Figure \ref{Fig0}). After $t=380$~s, the emission energy start increase again.During the tear-mode instability, a larger blob is formed and moves out. Now, after $t\sim400$~s, the magnetic flux structure is started to develop with a bigger size (compare to blob) in the simulation box.  This structure (see the last panel of both rows of Figure \ref{Fig0}) has generated  more non-thermal electrons, so we get a second broader peak after $t\sim450$~s (Figure \ref{Figbb}), although the light curve has a lot of rise and fall. The highest emission energy appears after $600$~s. Different  electron energy will lead to different radiation characteristics in different bands, so in the higher frequencies, the first energy peak looks smaller, the emission energy of the flux structure is higher and the simulated light curve look more difference than the simulated light curve of $230$~GHz and $43$~GHz.   In these two bands, the energy of the luminosity is around  $\sim10^{28}$~erg/s, which is higher than the radio band.\newline

Some numerical studies have shown that the emission of the synchrotron of the accretion flows and the jets on the SGR A*  can provide an excellent fit to the radio spectrum \citep{2000A&A...362..113F,2003ApJ...598..301Y}. For the synchrotron emission power-law, \citep{2003ApJ...598..301Y,2017MNRAS.468.2447P,2020A&A...638A...2G} have shown that the highest emission energy appears around the $\nu=10^{11}$~Hz to $\nu=10^{13}$~Hz as we have also found in our study (Figure\ref{Fig3}), which are plotted for two different times with the assuming three maximum energies of the electrons as mentioned in the each plot. For the other feature of the Lin-Forbis magnetic reconnection model \citep{2000JGR...105.2375L}, the flux structure begin to form after the blob ejected, and it takes hundreds of seconds to form well.  Figure \ref{Fig2} shows the second peak in $230$~GHz and $43$~GHz bands, the emission energy is starting to increase from $t\sim350$~s and then reach a high value after $t\sim400$~s. For the $\gamma_{1}=1$ and $\gamma_{2}=1000$ the total highest spectral energy has reached  $10^{27}$-$10^{28}$~erg/s, the highest energy appears near $\nu=10^{13}$~Hz. For the $\gamma_{1}=1$ and $\gamma_{2}=100$ the total highest spectral energy has reached  $10^{25}$-$10^{26}$~erg/s, the highest energy appears near $\nu=10^{12}$~Hz. And the highest emission energy lasts for a long period time until $t=650$~s. This process is the stage where the flux is formed and gradually develops, and the flux structure continues for $200$~s. The temperature in the flux structure is higher than that in the blob structures(see: \citet{2020MNRAS.499.1561Z}). It shows that this flux structure can also generate higher emission energy, and there is an obvious peak that can be seen in the observation. Compared with the theoretical spectrum of \citet{2003ApJ...598..301Y}, the magnetic strength is $20$~$G$,the $\gamma_{1}=1$ and $\gamma_{2}=10$ and they consider accretion flow, turbulence and magnetic reconnection events that occasionally accelerate a fraction of the electrons as the non-thermal synchrotron emission electrons, so the energy of emission is higher than our simulation. In the spectrum of Figure\ref{Fig3} it is only for the non-thermal synchrotron emission electrons in the blob or flux structure, and the simulation area is smaller, so the number  of the non-thermal synchrotron emission electrons is less than in \citet{2003ApJ...598..301Y} and the energy of the emission is lower. Because of the electron energy $\gamma$ is different, the highest energy on the synchrotron emission power-law will be the different frequency.  \newline

In the observations and the previous simulation works, the highest energy of the emissions can reach even more than $10^{31}-10^{35}$~erg/s. In  \citet{2020arXiv200514251B}, the strength of the magnetic field  is $35$~G, the number density of the electrons is $n_{0}=10^{6}$~$\mathrm{cm^{-3}}$, and the highest energy of the emissions is $10^{31}$~erg/s. In  \citet{2020MNRAS.497.4999D}, the highest energy of the emissions in the light curves in the wavebands at $3.8 $~$\mathrm{\mu m}$ and $2.2$~$\mathrm{\mu m}$ and the frequency at $230$~GHz during the flares can reach from $0.5 \times 10^{35}$~erg/s to $1.75 \times 10^{35}$~erg/s. The number density of the electrons is also $n_{0}=10^{6}$~$\mathrm{cm^{-3}}$. The strength of the magnetic field is $100$~G. The temperature of the electrons is $T_e \simeq 10^{12}$~K. The length scale of simulation in  \citet{2020arXiv200514251B,2020MNRAS.497.4999D} is much higher. The strength of the magnetic field in our simulation is $16$~G, the number density of the electrons is around $n_0 \approx10^{6}-10^{7}$~$\mathrm{cm^{-3}}$, the highest temperature is $T=3.8\times10^{10}$~K. Therefore, the energy of the emission is  $10^{25}$-$10^{26}$~erg/s in Radio band by synchrotron emission, and $10^{27}$-$10^{28}$~erg/s in IR band by the non-thermal synchrotron emission, which are much less than the observed one and the previous simulations with a large area. Thus, we are expecting that with a larger area of simulation,  the energy of the emissions will be higher. Moreover, the trend of the light-curve at $230$~GHz and $43$~GHz is not very similar to the study in  \citet{2020arXiv200514251B,2020MNRAS.497.4999D}, because the formation of the flux structure continues for a long time, and it does not immediately rise and fall to form a standard peak pattern. For all these, the energy of the emission of the single blob can reach $10^{26}$~erg/s, so that it can be observed. In reality,  it is possible that a single flare consists of multiple blobs or a group of flares so that the observations can get higher luminosities. On the other hand, the formations of the blob and the flux structure, which are the two processes of the magnetic reconnection, can also explain that why some flares have double-peaks in the observations while others with the single process only have a single peak \citep{2020arXiv200514251B}. \newline

\section{Conclusion}
\label{sec4}
This work is based on the numerical studies of the magnetic reconnection during the blob formation and ejection from the radiatively inefficient accretion flow around the massive black hole \citep{2020MNRAS.499.1561Z}. The main conclusions from our numerical simulations are as follows:\newline

1. From the energy of the emission via time at $230$~GHz and $43$~GHz, there are two peaks. The one peak is due to blob formation, and the second peak is due to magnetic flux structure formation, which is much persistent than the first process in Figure \ref{Fig2}. 
The second peak is formed after the tear-mode instability with ejecting of the blob from the simulation area.  
The non-thermal synchrotron emission energy reaches 
$10^{25}$-$10^{26}$~erg/s for two frequencies, and this can be observable of the single blob or flux structure. The double peak light curve with two different processes can be one of the explanations of the observed double peaks \citep{2020arXiv200901859G,2021ApJ...917....8B}. For the non-thermal synchrotron emission at $3.8$~$\mathrm{\mu m}$ and $2.2$~$\mathrm{\mu m}$ reaches $10^{27}$-$10^{28}$~erg/s for the single blob and flux structure. 
So we infer that the observed high emission ($\sim10^{31-35}$ erg/s) in radio and  infrared flares can be due to a group of flares or blobs/flux structures in the large simulation area of the disk.
\newline 

2. The spectral for the synchrotron radiation show the highest emission energy appear near $\nu=10^{11}$~Hz for the $\gamma_{1}=1$ and $\gamma_{2}=100$ and $\nu=10^{13}$~Hz, for the $\gamma_{1}=1$ and $\gamma_{2}=1000$, which is the same as the synchrotron emission power-law in the previous work such as \citep{2003ApJ...598..301Y,2017MNRAS.468.2447P,2020A&A...638A...2G}, that is the peak value appears between $\nu=10^{12}$~Hz-$\nu=10^{13}$~Hz . So we believe that the highest energy on the synchrotron emission power-law is peak at the different frequency because of the electron energy $\gamma$ is different during the blob or flux structure evolution. \newline

Regarding the prediction of the flare model observations on the accretion disk of the black hole, we infer based on the solar coronal jets, flares, and other phenomena. The ejection of the solar will be observed on a multi-waveband, they're very close to us, so we can see them at a clearer resolution. If the flares in the inner area of the accretion disk are strong enough and the size is large enough, the formation of blobs or flares will be easily observed. Different wavebands represent different levels of information in the solar atmosphere, and some activity phenomena are multi waveband responses, and they can also be observed in multiple wavebands, such as radio, NIR \citep{2018A&A...618L..10G, 2020ApJ...891L..36G}, X-ray \citep{2017MNRAS.468.2447P}. For the inclination angle of observer, the SGRA* flares observed by \citet{2018A&A...618L..10G} show the inclination angle is $160^{\circ}$. However this is not the fixed value, many other simulation work\citet{2021PASJ...73..912T,2021MNRAS.504.6076R} set the different inclination angle, such as \citet{2021ApJ...917....8B} set the inclination angle is $180^{\circ}$. In order to simplify the calculation process, we set the inclination angle is $180^{\circ}$, that is, the line of sight is perpendicular to the accretion disk and has no Doppler effect. If the inclination angle changes, the radiation energy increases due to Doppler effect and decreases due to the shielding of the disk.   On the other hand, because of  the effect of space-time bending on light is not considered, ray-tracing\citep{2020ApJ...897..148G,2013ApJ...777...13C,2016MNRAS.462..115D} is not used in this calculate of the emission.  Therefore, our future simulations will be conducted in stronger magnetic fields and larger simulation areas.  We will also try to predict radiation energy in multiple bands. And we will consider more fine calculation conditions such as the inclination angle of observer, the ray-tracing and so on.\newline

\section*{Acknowledgements}
We would like to thank the Prof. YeFei Yuan, Prof. Yanrong Li, Prof. Feng Yuan, Prof. Lei Ni ,Prof Liang Chen,  Doctor Xi Lin and Doctor Yinhao Wu, Doctor Shanshan Zhao for their helpful comments. 
This work is supported by National Natural Science Foundation of China (Grant Nos. 11725312, 11421303).
The numerical calculations in this paper have been done on supercomputing system in the Super-computing Center of University of Science and Technology of China. The figures in this paper are plotted on Super-computing in Prof. Wang,  Jun Xian's group.

\section*{Data Availability}

The data underlying this article will be shared on reasonable request to the corresponding author.




\bibliographystyle{mnras}
\bibliography{blobformation}

\newpage

\begin{figure}
	\centering
	
	\includegraphics[width=0.8\textwidth, angle=0]{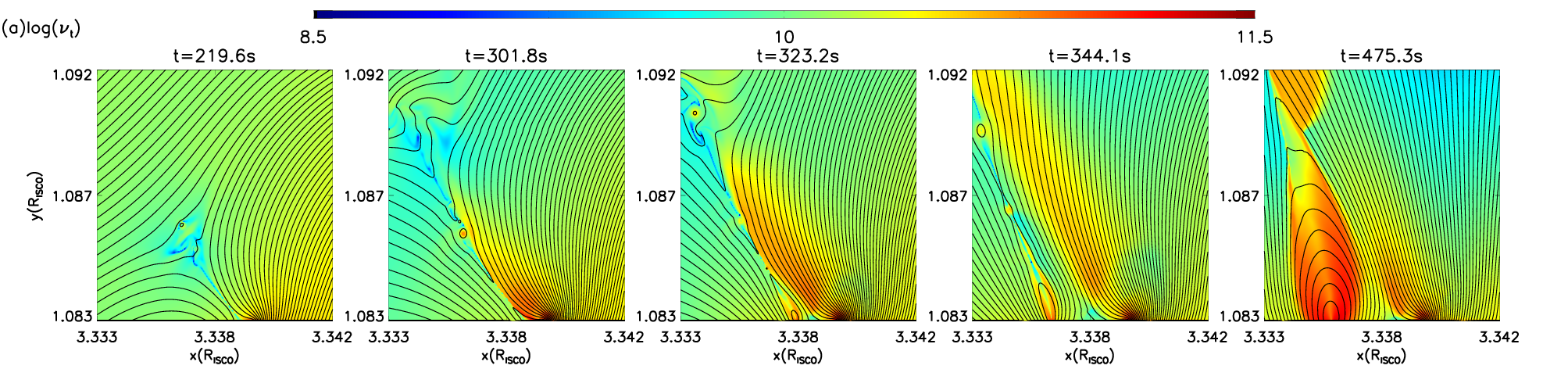}
	
	\caption{Distribution of the turnover frequency ($\nu_t$) of the plasma.}
	\label{Fign0}
\end{figure}

\begin{figure}
	\centering
	\begin{minipage}[t]{0.48\textwidth}
		\centering
		\includegraphics[width=0.8\textwidth, angle=0]{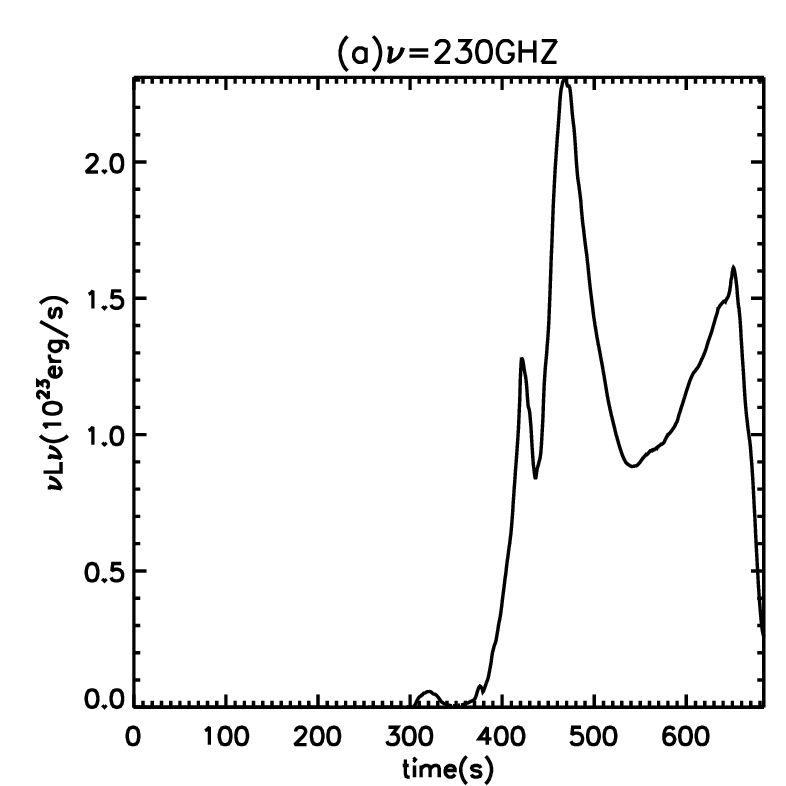}
	\end{minipage}
	\caption{Distributions of thermal synchrotron emission energy change via time at band $230$~$GHz$.}
	\label{Fig00}
\end{figure}

\begin{figure}
	\centering
	\includegraphics[width=0.8\textwidth, angle=0]{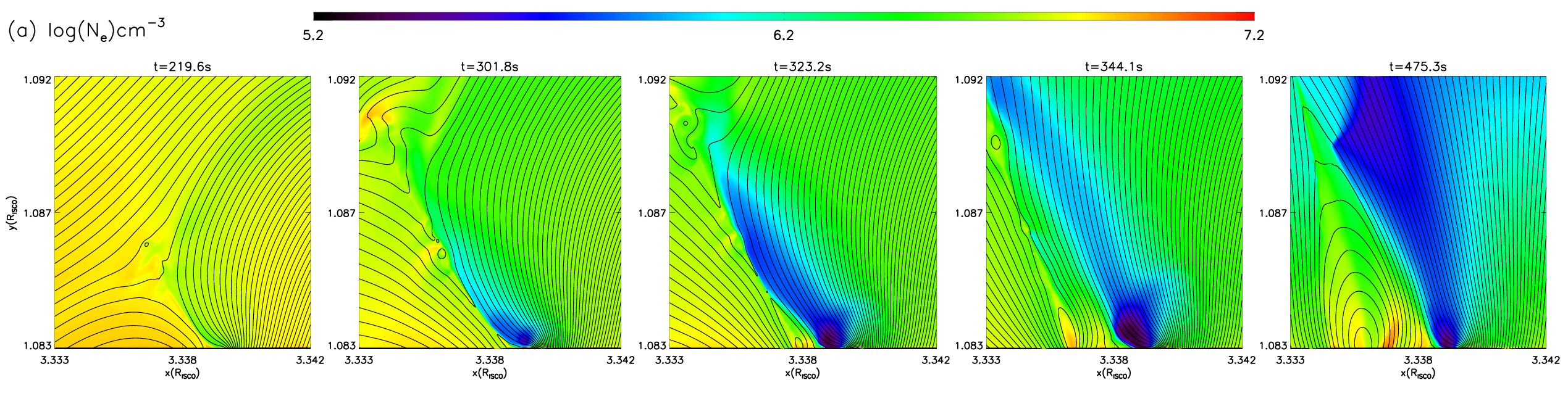}
	\includegraphics[width=0.8\textwidth, angle=0]{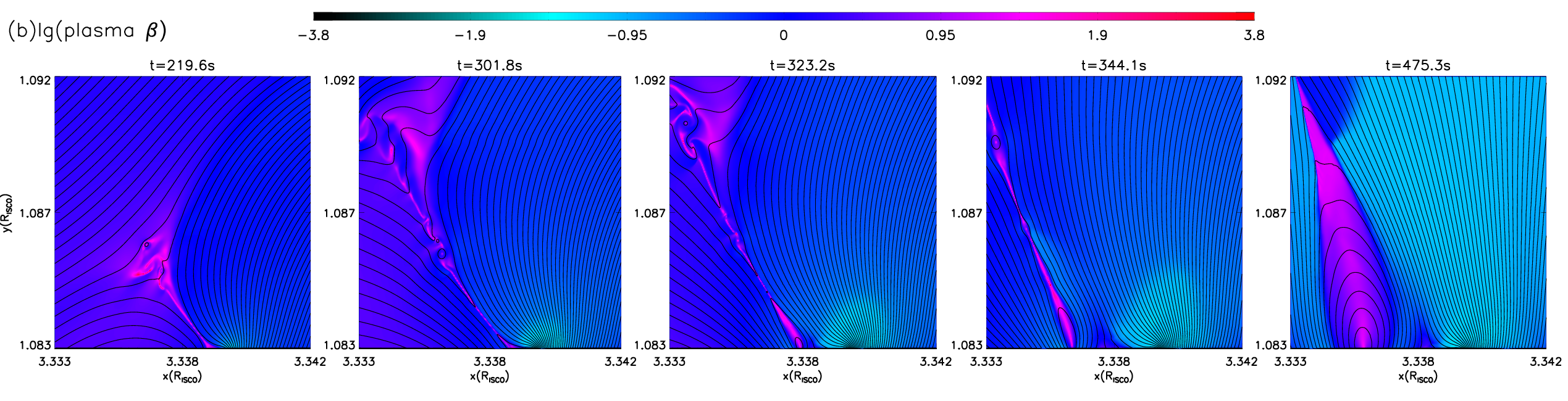}
	\caption{Distribution of the electron number density, and $\beta$ are presented in upper and lower row with sequence of five instantaneous times, respectively.}
	\label{Fig0}
\end{figure}

\begin{figure}
	\centering
	\begin{minipage}[t]{0.48\textwidth}
		\centering
		\includegraphics[width=0.8\textwidth, angle=0]{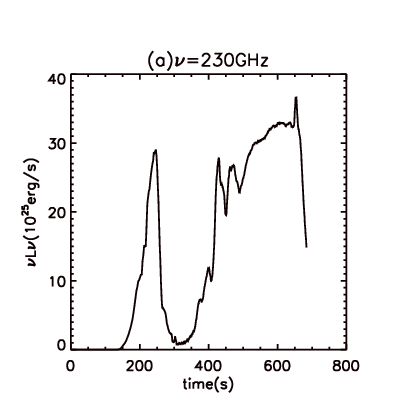}
	\end{minipage}
	\begin{minipage}[t]{0.48\textwidth}
		\centering
		\includegraphics[width=0.8\textwidth, angle=0]{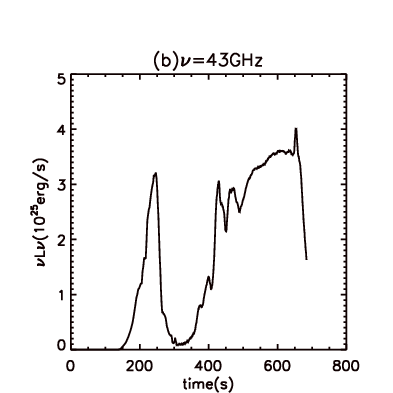}
	\end{minipage}
	\caption{Distributions of non-thermal synchrotron emission energy change via time at different band (a) $230$~$GHz$ (b) $43$~$GHz$.}
	\label{Fig2}
\end{figure}

\begin{figure}
	\centering
	\begin{minipage}[t]{0.48\textwidth}
		\centering
		\includegraphics[width=0.8\textwidth, angle=0]{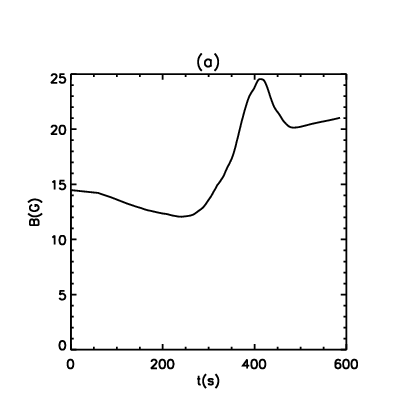}
	\end{minipage}
	
	\caption{Distributions of average magnetic field strength in the simulation area $B$ change via time.}
	\label{Fig1}
\end{figure}

\begin{figure}
	\centering
	\begin{minipage}[t]{0.48\textwidth}
		\centering
		\includegraphics[width=0.8\textwidth, angle=0]{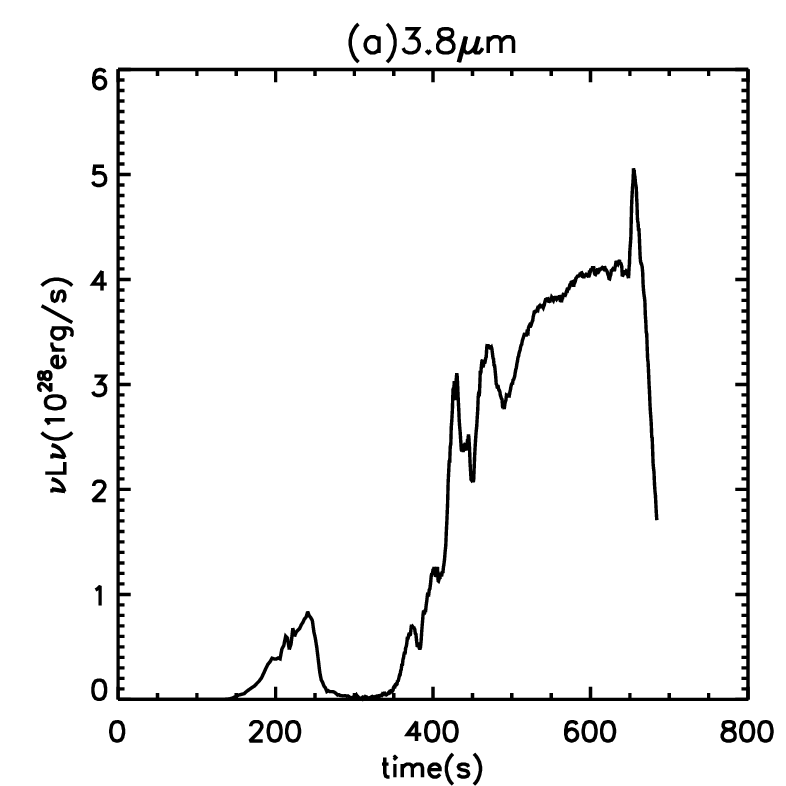}
	\end{minipage}
	\begin{minipage}[t]{0.48\textwidth}
		\centering
		\includegraphics[width=0.8\textwidth, angle=0]{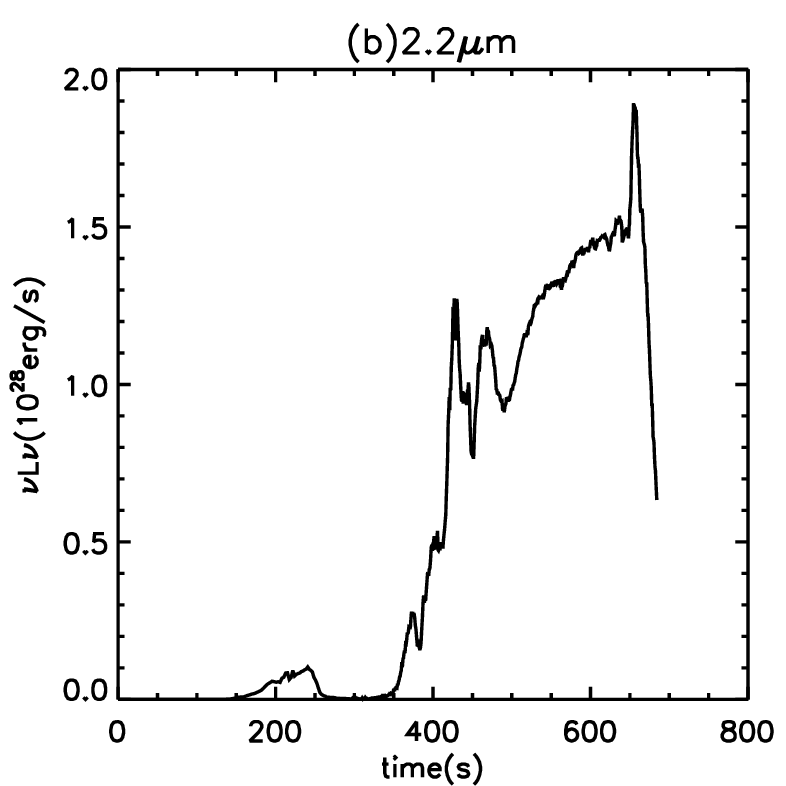}
	\end{minipage}
	\caption{Distributions of non-thermal synchrotron emission energy change via time at different band (a) 3.8 $\mu m$ (b) 2.2 $\mu m$.}
	\label{Figbb}
\end{figure}

\begin{figure}
	\centering
	\begin{minipage}[t]{0.48\textwidth}
		\centering
		\includegraphics[width=0.8\textwidth, angle=0]{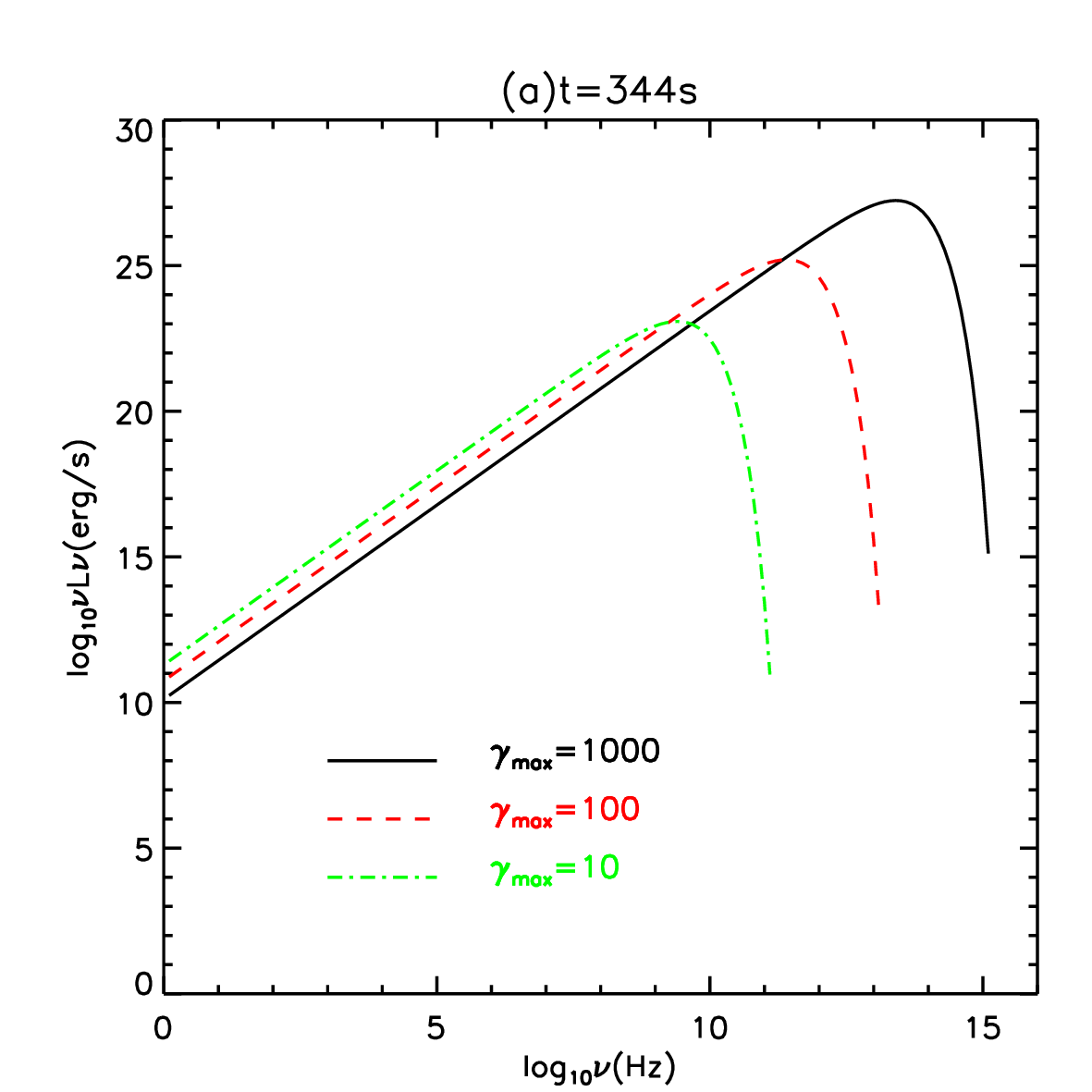}
	\end{minipage}
	\begin{minipage}[t]{0.48\textwidth}
		\centering
		\includegraphics[width=0.8\textwidth, angle=0]{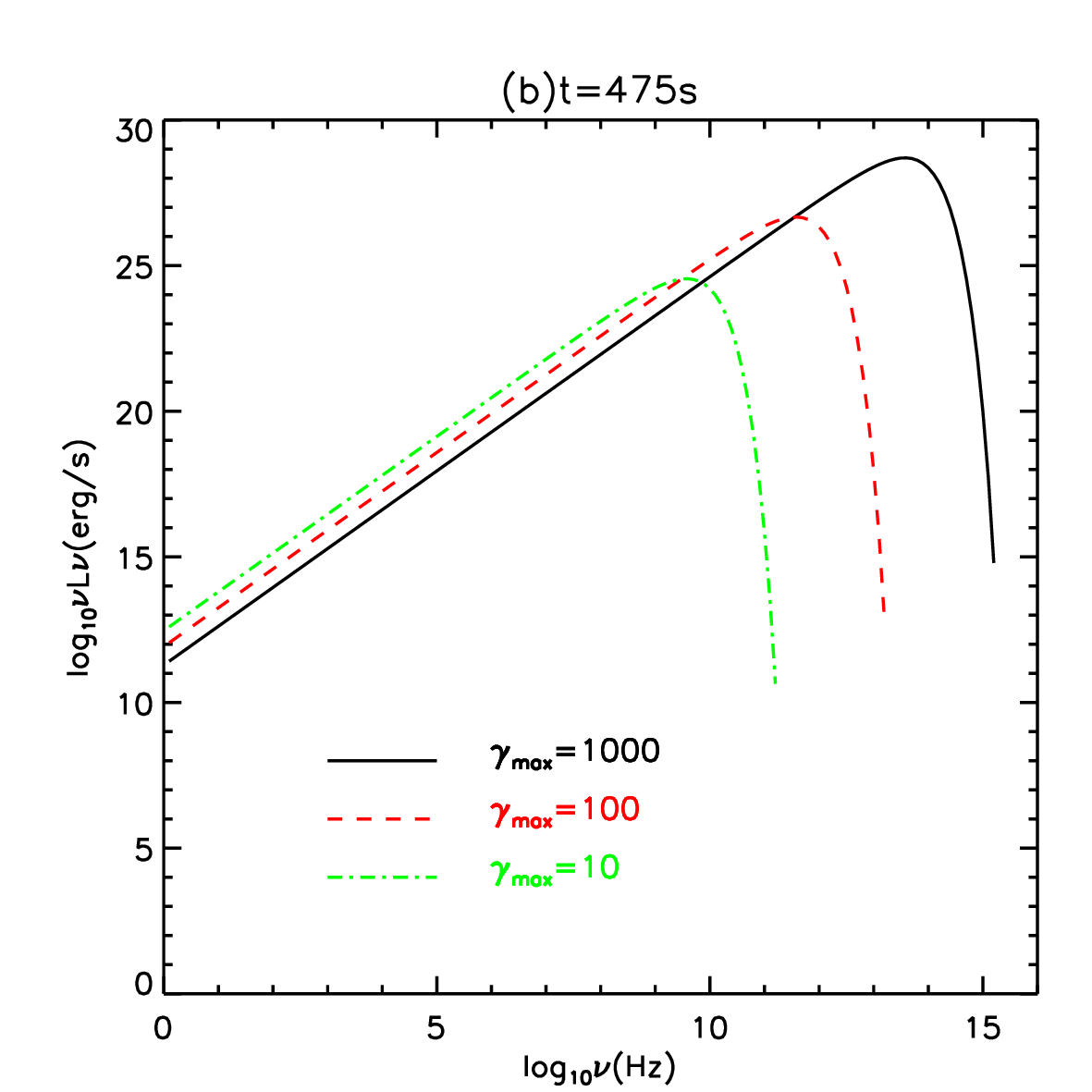}
	\end{minipage}
	\caption{Spectral for the synchrotron radiation at different band for different $\gamma$. (a) For the blob structure. (b) For the flux structure.}
	\label{Fig3}
\end{figure}



\bsp	
\label{lastpage}
\end{document}